\documentclass{article}
\usepackage{graphicx} 
\usepackage{amsmath}
\usepackage{amsfonts}
\usepackage{amssymb}
\usepackage{float}
\usepackage{url}
\usepackage[toc,page]{appendix}
\usepackage{tabularx}
\usepackage[margin=1in]{geometry}
\usepackage{color}
\usepackage[normalem]{ulem}
\usepackage{authblk}

\title{Not All Accuracy Is Equal: Prioritizing Independence in Infectious Disease Forecasting}
\author[1, 2]{Carson Dudley\footnote{Corresponding author, cdud@umich.edu}}
\author[1, 2]{Marisa Eisenberg}
\affil[1]{Department of Mathematics, University of Michigan, 530 Church Street, Ann Arbor, MI 48109, USA}
\affil[2]{School of Public Health, University of Michigan, 1415 Washington Heights, Ann Arbor, MI 48109, USA}
\date{\today}

\begin{document}

\maketitle

\section*{Abstract}
    Ensemble forecasts have become a cornerstone of large-scale disease response, underpinning decision making at agencies such as the US Centers for Disease Control and Prevention (CDC). Their growing use reflects the goal of combining multiple models to improve accuracy and stability versus relying on any single model. However, while ensembles regularly demonstrate stability against individual model failures, improved accuracy is not guaranteed. During the COVID-19 pandemic, the CDC's multi-model ensemble outperformed the best single model by only 1\%, and CDC flu ensembles have often ranked below individual models.
    
    Prior work has established that ensemble performance depends critically on diversity: when models make independent errors, combining them yields substantial gains. In practice, however, this diversity is often lacking. Here, we propose that this is due in part to how models are developed and selected: both modelers and ensemble builders optimize for stand-alone accuracy rather than ensemble contribution, and most epidemic forecasts are built from a small set of approaches trained on the same surveillance data. The result is highly correlated errors, limiting the benefit of ensembling.
    
    This suggests that in developing models and ensembles, we should prioritize models that contribute complementary information rather than replicating existing approaches. We present a toy example illustrating the theoretical cost of correlated errors, analyze correlations among COVID-19 forecasting models, and propose improvements to model fitting and ensemble construction that foster genuine diversity. Ensembles built with this principle in mind produce forecasts that are more robust and more valuable for epidemic preparedness and response.

\section*{Introduction}

Ensemble forecasts have rapidly become central to epidemic response and preparedness \cite{ensembling, ReichRay2022Collaborative}. In the United States, the Centers for Disease Control and Prevention (CDC) have coordinated multi-model ensembles for COVID-19 through the COVID-19 Forecast Hub \cite{evaluationpnas}, for influenza through the long-running FluSight challenge \cite{flusight}, and most recently for RSV \cite{us_rsv_forecast_hub}. A new metro-scale hub is now in development to deliver local-level ensemble forecasting for respiratory diseases \cite{flu_metrocast_hub}. The proliferation of forecasting ensembles reflects a sound rationale: before forecasts are submitted, there is no way to know which individual model will perform best, and combining multiple models reduces the risk of relying on any single one that might fail. Moreover, when contributors provide complementary information, ensembles can exceed the accuracy of all individual models, making them a valuable tool for public health decision-making.

Yet while infectious disease forecasting ensembles have demonstrated stability and consistency against individual model failure \cite{ray2018prediction}, they have often underperformed in improving accuracy beyond their component models. For COVID-19, the CDC’s multi-model ensemble from April 2020 through November 2021 was only 1\% more accurate than the single best model, despite dozens of contributors \cite{evaluationpnas}. For influenza, the CDC’s FluSight ensemble did not lead the field: in 2021 it ranked second overall, and in 2022 it ranked fifth out of 17 models \cite{flusight}. These results indicate that current ensembles frequently provide only marginal gains in accuracy compared to individual models.

The importance of diversity in ensemble methods is well established in the broader machine learning literature. Theory and practice have shown that ensembles benefit most when component models make independent errors, and a variety of metrics have been proposed to quantify and encourage this diversity \cite{fan2025diverse, div, Wu_2021_CVPR}. Despite this recognition, infectious disease forecasting ensembles often have highly correlated errors.

Why does this occur? Both forecasting groups and ensemble designers typically optimize for individual model performance: models are tuned to minimize their own error, and ensemble weights are often assigned according to those same error scores \cite{weighting, Fox2024Optimizing}. By primarily rewarding stand-alone accuracy, current practice tends to favor models that resemble one another, reducing rather than increasing diversity between models. In epidemic forecasting this problem is amplified because most submissions come from a narrow set of approaches: mechanistic frameworks (e.g., compartmental or agent-based models) \cite{desikan2022mechanistic, ihme2021modeling}, statistical time-series approaches \cite{statistical}, or machine learning techniques like neural networks \cite{deepcovid}. While many contemporary models are hybrids of these approaches \cite{deepgleam, rodriguez2022einns, mantis}, they nonetheless rely on the same fundamental mathematical structures and data assumptions, resulting in some methodological redundancy. Across these methods, modelers also often necessarily rely on a similarly limited pool of covariates: for COVID-19, case counts, hospitalizations, and deaths were the dominant predictors \cite{evaluationpnas}, with only a few groups incorporating additional information such as mobility, wastewater, or internet search data \cite{deepcovid, SantillanaDigitalTraces}. The result is that many models in an ensemble, despite surface-level differences, depend on the same underlying methods and data, and thus tend to make correlated errors.

\begin{figure}[ht!]
  \centering
  \includegraphics[width=0.97\linewidth]{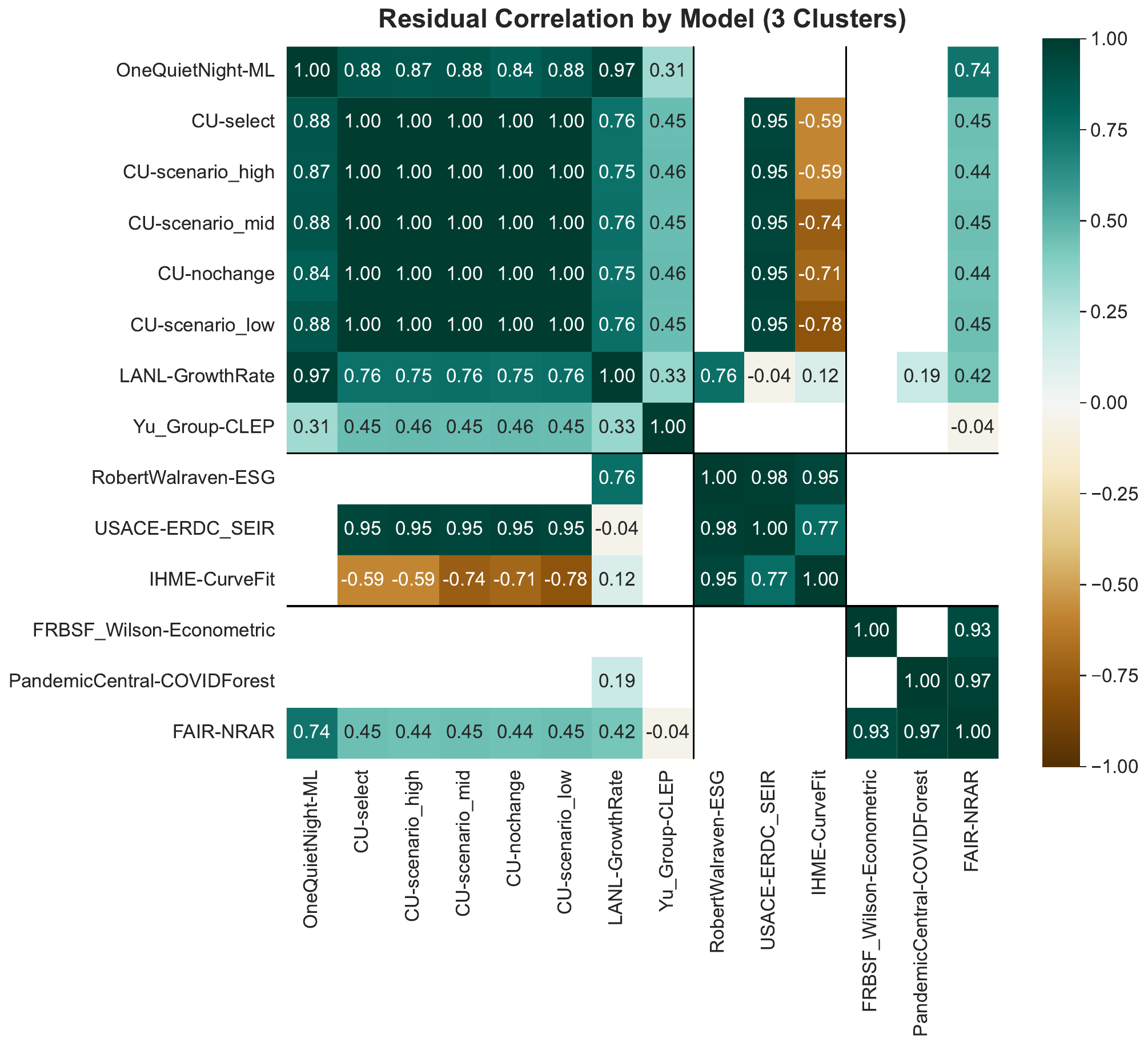}
  \caption{\textbf{Residual correlation structure among forecasting models.}
Heatmap of Pearson correlations between model residuals for selected case forecasting models in the CDC COVID-19 Forecast Hub from July 2020 to December 2022. Models were restricted to those with at least one year of overlapping forecasts with at least one other model for weekly cases. Models are clustered into three groups using agglomerative clustering. Cell annotations report correlation values. Extremely high within-cluster correlations (often $>0.95$) contrasted with much weaker or negative between-cluster correlations indicate that the ensemble was composed of only a few distinct families of models, limiting potential gains when diversity is low. White squares indicate that two models never had a full year of common overlap. Apparent discrepancies (e.g., IHME correlating positively with USACE but negatively with CU models, even though USACE and CU are strongly correlated) arise because pairs were evaluated on different periods of overlap.
}

  \label{fig:corr_heatmap_n3}
\end{figure}

To quantify the correlation structure among case forecasting models submitted to the CDC's COVID-19 Forecast Hub, we computed pairwise Pearson correlations of residuals (forecast minus observed values) across a subset of models from the hub\footnote{We filtered to only models that submitted all 1-week incident case forecasts at least seven days ahead. If we relax the filter to models submitting 1-week cases less than seven days ahead, there are more eligible models. However, even when the number of eligible models gets up to 58 and we allow 20 clusters, the first cluster has 14 models in it with an average correlation of $0.86.$ Code is available at \texttt{github.com/carsondudley1/ensemble\_analysis}.} \cite{forecasthub}. The resulting correlation matrix was clustered to identify groups of similar models. Without clustering, the average correlation was approximately $0.60$. When grouped into three clusters, the within-cluster correlation rose to $0.82$, and with four clusters it reached $0.91$. Figure~\ref{fig:corr_heatmap_n3} visualizes this structure, showing that many models exhibit extremely high correlations ($r>0.95$). These patterns demonstrate that the ensemble was effectively composed of only a few distinct families of models, rather than dozens of independent perspectives.

When most models make similar mistakes, adding more of them to an ensemble yields little benefit. As a result, ensembles are often only marginally better than their component parts---or, in some cases, worse. To overcome this limitation, we must focus on deliberately fostering diversity and complementarity among contributors.

\section*{An illustrative toy example: why ensembles underperform}

The central issue is correlation: when models fail in similar ways, ensembling cannot deliver its theoretical benefits. A simple quantitative argument illustrates how stark the gap can be between the ideal case and what we observe in practice.

We consider the mean absolute error of a model relative to a naïve persistence baseline, which simply projects the most recent observation forward. Normalizing by a baseline provides a dimensionless measure of skill, allowing us to compare and aggregate model performance across different locations and epidemic scales where absolute error magnitudes vary significantly. If the baseline error is $E_0$ and a model has error $E$, then
\[
\text{relative MAE} = \frac{E}{E_0}.
\]
A value of $0.9$ means the model’s error is $90\%$ of the baseline, a $10\%$ improvement. Lower values therefore indicate better forecasts.

Now consider an ensemble of $N$ models, each with the same individual relative MAE. If the models are unbiased estimators of the true values and the errors of the models are uncorrelated, simple averaging reduces the error variance by a factor of $N$. In this ideal case the ensemble error is
\[
\text{relative MAE}_{\mathrm{ens}} = \frac{\text{relative MAE}_{\mathrm{model}}}{N}.
\]

In the COVID-19 Forecast Hub, there were $N=28$ contributing models between April 2020 and November 2021 \cite{evaluationpnas}. The median model had relative MAE of $0.87$, corresponding to a $13\%$ improvement over the baseline \cite{evaluationpnas}. The ensemble achieved $0.66$ relative MAE ($34\%$ improvement), only slightly better than the best single model at $0.67$.

To illustrate the benefit of ensembling under ideal conditions, consider a scenario where every contributor was much weaker, with $\text{relative MAE}=0.95$ (a $5\%$ improvement over baseline). If their errors were uncorrelated and the models were unbiased, the ensemble would have had
\[
\text{relative MAE}_{\mathrm{ens}} \approx \frac{0.95}{28} \approx 0.03,
\]
corresponding to a $97\%$ improvement over the baseline. Even weaker individual models would combine into an exceptionally strong ensemble under these (relatively strong) assumptions.

\begin{figure}[ht!]
    \centering
    \includegraphics[width=0.7\textwidth]{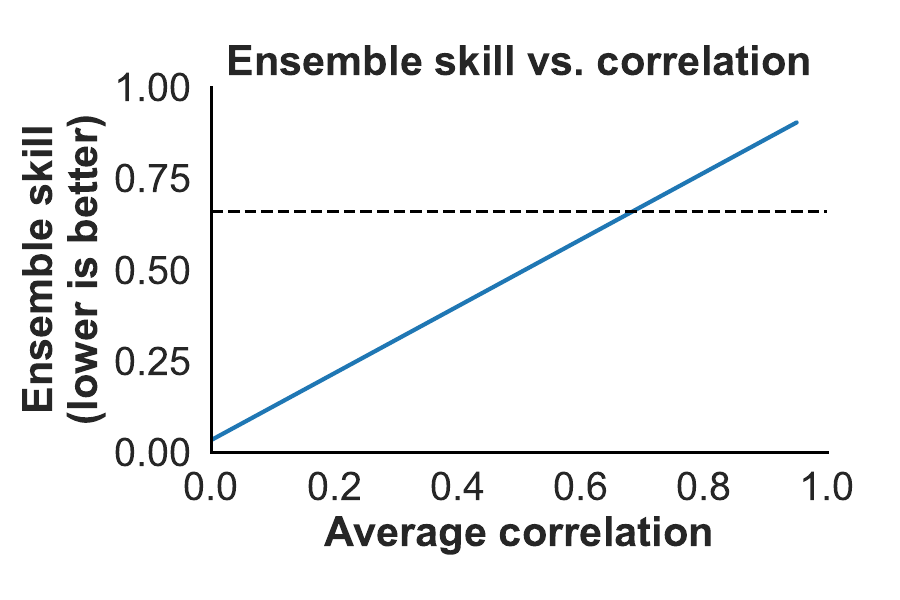}
    \caption{\textbf{Impact of error correlation on ensemble performance.} 
    The blue line shows the expected ensemble skill for $N=28$ unbiased models of equal quality, each only 5\% better than the baseline (versus 13\% median skill from the COVID-19 ensemble \cite{evaluationpnas}). 
    As correlation increases, the benefit of ensembling declines sharply. 
    The dashed line marks the observed performance of the CDC COVID-19 ensemble ($\approx 0.66$) \cite{evaluationpnas}, far above the theoretical potential ($\approx 0.03$ if errors were uncorrelated). 
    This gap reflects the high correlation of errors among current models, limiting ensemble gains.}
    \label{fig:ensemble_correlation}
\end{figure}

The gap between the theoretical $97\%$ improvement and the observed $34\%$ improvement reflects the fact that contributors’ errors were not independent but highly correlated (as well as that most models were likely at least somewhat biased estimators). With moderate correlation, the expected improvement drops to around $50\%$; with strong correlation, to the $20\%$s. This relationship is illustrated in Figure~\ref{fig:ensemble_correlation}, which plots the expected ensemble error as average correlation increases, assuming 28 models of equal skill whose pairwise error correlations are all equal to the value on the x-axis. This simple example shows how even modestly accurate models could combine into a near-perfect ensemble under independence, but also how quickly those gains erode as error correlation increases.

\section*{A contribution-focused framework for ensemble design}

We have seen that stand-alone accuracy, while important, is not sufficient to guarantee that ensembles can reach their full potential. To reach the accuracy gains that are possible for ensembles, we must also assess whether each model contributes complementary information. A useful way to frame this is to decompose a model’s performance into two parts: the component of its errors that are correlated with the current ensemble’s errors, and the component that are independent. The correlated component reinforces patterns the ensemble already captures. The independent component provides new information that can substantially improve the ensemble’s accuracy and resilience. This independent signal must still outperform naïve baselines such as projecting the most recent observation forward. Within that boundary, however, a model that adds even a modest amount of novel information can be more valuable than one that achieves higher stand-alone accuracy but contributes only redundant signals. A key aspect of ensemble design should therefore be to maximize this independent component while ensuring that all models improve on trivial forecasts.

This perspective has different implications for modelers and ensemblers. For modelers, the goal is not only to minimize stand-alone forecasting error, but to design models that contribute information not already captured by others. This can mean drawing on new data streams or adopting alternative model structures that introduce fundamentally different assumptions or learning mechanisms (e.g., novel methods such as diffusion-based approaches \cite{influpaint}, epimodulation \cite{epimodulation}, or simulation-grounded neural networks \cite{sgnns, sgnntheory}). Training objectives can be broadened beyond error minimization: models could be trained to penalize correlation with ensemble residuals, ensuring they capture independent signal, or to emphasize alternative performance criteria. For instance, an objective might reward accurate detection of turning points in epidemic curves, or weight errors more heavily when they would alter a public health decision (e.g., underestimating a surge in hospital demand). Even if these models are less accurate on traditional metrics, they may add more value by reducing shared errors within the ensemble and by highlighting aspects of the forecast most relevant for decision-making.

For ensemblers, this principle implies weighting schemes that account for both accuracy and correlation, giving greater influence to models that diversify ensemble errors. Ensemblers can also design sub-ensembles that cluster similar models together before combining them, reducing the penalty of correlated errors and producing forecasts that are more robust across epidemic regimes.

Other fields already adopt this principle. In climate and weather forecasting, for example, the design of multi-model ensembles explicitly emphasizes methodological diversity, recognizing that skill arises from combining models that fail in different ways \cite{weather}. Infectious disease forecasting should adopt the same approach: moving beyond size for its own sake to intentionally cultivating diversity and complementarity.

\section*{Practical implications for disease forecasting}

Translating this framework into practice requires shifts at several levels of the forecasting ecosystem. For forecast hubs and public health agencies, one useful step would be to publish not only the stand-alone skill of each model but also the improvement it provides beyond the ensemble baseline. Others have called for evaluation to emphasize marginal contribution rather than focusing solely on leaderboard rankings \cite{beyondleaderboards}, reflecting a growing recognition that independent value matters in addition to accuracy. Additional reporting, such as error correlation matrices or regime-specific evaluations, would further help identify clusters of similar models and point to opportunities for broadening methodological diversity.

Hubs can also shape incentives. Current challenges largely report and reward stand-alone accuracy \cite{evaluationpnas, flusight}, but new metrics could recognize contributions that increase ensemble diversity. Explicitly valuing complementary information would encourage submissions built on novel data sources or alternative modeling approaches, rather than incremental improvements of familiar designs.

\section*{Conclusion}

Ensemble forecasts have become central to epidemic preparedness, but when all models in an ensemble rely on similar methods and data, their errors are highly correlated, and the benefits of ensembling are reduced. To address this, we must value independent contributions---aspects of a model that provide information beyond what the ensemble already captures. 

Doing so presents some practical challenges. For example, correlations between models can shift over time and across targets, and biases shared across models cannot be averaged away. Yet these issues can be managed: correlations can be monitored in rolling windows, and diversification helps reduce shared biases. Transparent reporting of both stand-alone and independent skill would provide a clearer picture of each model’s contribution and help guide future innovation.

Ensembles achieve their value not only by aggregating models, but by intentionally combining models that contribute complementary information. For modelers, this means that innovation in data, methods, and objectives should be guided not only by accuracy against baselines but by the potential to add independent value to ensembles. For ensemblers, this means adopting weighting and evaluation strategies that recognize contribution as well as accuracy. By embracing these principles, the forecasting community can build ensembles that are more diverse, more resilient, and ultimately more useful for epidemic preparedness and response.

\section*{Acknowledgements}

This publication was made possible by the Insight Net cooperative agreement with University of Michigan (5 NU38FT000002-02-00) from the CDC’s Center for Forecasting and Outbreak Analytics (CDC-RFA-FT-23-0069). Its contents are solely the responsibility of the authors and do not necessarily represent the official views of the Centers for Disease Control and Prevention.

\bibliographystyle{unsrt}

\end{document}